\documentclass[reprint,amsmath,amssymb,superscriptaddress,aps,prb]{revtex4-1}
\usepackage[utf8x]{inputenc}
\usepackage{graphicx}
\usepackage{dcolumn}
\usepackage{bm}
\usepackage{soul}
\usepackage{subfigure}
\usepackage{multirow}
\usepackage{hhline}
\usepackage{siunitx}
\usepackage[usenames,dvipsnames]{color}

\begin{document}

\preprint{APS/123-QED}

\title{Impurity-dependent gyrotropic motion, deflection and pinning of current-driven ultrasmall skyrmions in PdFe/Ir(111) surface}

\author{Imara  Lima Fernandes}
\email{i.lima.fernandes@fz-juelich.de}
\affiliation{Peter Gr\"unberg Institut and Institute of Advanced Simulation, Forschungszentrum J\"ulich \& JARA, 52425 J\"ulich, Germany}

\author{Jonathan Chico}
\affiliation{Peter Gr\"unberg Institut and Institute of Advanced Simulation, Forschungszentrum J\"ulich \& JARA, 52425 J\"ulich, Germany}

\author{Samir Lounis}
\email{s.lounis@fz-juelich.de}
 \affiliation{Peter Gr\"unberg Institut and Institute of Advanced Simulation, Forschungszentrum J\"ulich \& JARA, 52425 J\"ulich, Germany}

\date{\today}

\begin{abstract}
Resting on multi-scale modelling simulations, we explore dynamical aspects characterizing magnetic skyrmions driven by spin-transfer-torque towards repulsive and pinning 3$d$ and 4$d$ single atomic defects embedded in a Pd layer deposited on the Fe/Ir(111) surface. The latter is known to host sub-10 nm skyrmions which are of great interest in information technology.  The Landau-Lifshitz-Gilbert equation is parametrized with magnetic exchange interactions extracted from the ab-initio all-electron full potential Korringa-Kohn-Rostoker Green function method, where spin-orbit coupling is added self-consistently. Depending on the nature of the defect and the magnitude of the applied magnetic field, the skyrmion deforms by either shrinking or increasing in size, experiencing thereby elliptical distortions. After applying a magnetic field of 10 Tesla, ultrasmall skyrmions are driven along a straight line towards the various defects which permits a simple analysis of the impact of the impurities. Independently from the nature of the skyrmion-defect complex interaction, being repulsive or pinning, a gyrotropic motion is observed. A repulsive force leads to a skyrmion trajectory similar to the one induced by an attractive one. We unveil that the circular motion is clockwise around pinning impurities but counter clockwise around the repulsive ones, which can be used to identify the interaction nature of the defects by observing the skyrmions trajectories. Moreover, and as expected, the skyrmion always escapes the repulsive defects in contrast to the pinning defects, which require a minimal depinning current to observe impurity avoidance. This unveils the richness of the motion regimes of skyrmions. We discuss the results of the simulations in terms of the Thiele equation, which provides a reasonable qualitative description of the observed phenomena. Finally, we show an example of a double track made of pinning impurities, where the engineering of their mutual distance allows to control the skyrmion motion with enhanced velocity.
\end{abstract}

\maketitle

\section{Introduction}
In the last decades a significant effort has been devoted to the exploration of alternatives to \textit{electronics} for information technology applications. One of the most prominent, the field of \textit{spintronics}, makes use of both the charge and the spin degree of freedom of the electron to transmit and store information. An appealing prospected device, the so-called racetrack memory~\cite{Parkin_racetrack}, has recently emerged. It is based on dynamical magnetic bits, where the information is encoded in magnetic domain walls which are displaced by using spin polarized currents, due to the spin transfer torque (STT)~\cite{Slonczewski,Berger1974,Berger1984,Berger1986}. However, applications based on domain walls suffer from a key weakness, that is they become easily pinned on defects present in the materials. To overcome the difficulties steaming from the presence of defects, a different kind of magnetic texture has been proposed recently as a way to transmit information in racetracks, namely \textit{skyrmions}~\cite{Uchida359,yu2010real}.

Skyrmions are particle like non-collinear magnetic textures~\cite{Bogdanov,Roessler2006}, whose topological character~\cite{nagaosa_topological_2013} enhances their stability against external perturbations giving rise to various interesting properties\cite{Wiesendanger2016,fert2017advances,Dias2016}. These textures are often the result of a competition between the Heisenberg exchange and the relativistic Dzyaloshinskii-Moriya interaction (DMI)~\cite{Dzyaloshinsky1958241,PhysRev.120.91}, which is present in materials lacking inversion symmetry and requesting a finite spin orbit coupling. Although initially found in non-centrosymmetric B20 materials ~\cite{Uchida359,yu2010real,PhysRevB.77.184402,yu2011near}, they have been also found in ultra-thin magnetic layers deposited in heavy metals~\cite{dupe_tailoring_2014,jiang_blowing_2015,heinze_spontaneous_2011} due to the large interfacial DMI~\cite{PhysRevLett.44.1538,Crepieux1998341,PhysRevB.90.054412}.

The allure of skyrmions for spintronic applications lies on the fact that the current thresholds needed to achieve skyrmionic motion is orders of magnitude smaller than the ones needed for domain walls~\cite{Yu2012,fert_skyrmions_2013}. Another advantage of skyrmions predicted from phenomenological based models is their capacity to avoid defects~\cite{iwasaki_universal_2013,fert_skyrmions_2013}, that is in wide nanowires, skyrmions are able to move around defects due to their particle like nature, unlike what occurs for magnetic domain walls. However, recent work has shown that in constrained geometries, such as in nanowires, defects can substantially affect the motion of skyrmions~\cite{iwasaki_current-induced_2013,hanneken_pinning_2016,stosic_pinning_2017}. 

Various experiments of interface stabilized skyrmions in ultrathin heavy metal/ferromagnetic bilayers and multilayers unravelled a  complex defect-driven dynamical behavior as function of applied currents~\cite{Woo2016,Legrand2017,Jiang2016,Litzius2017}. Magnetic vortices with cores containing thousands of atoms can be driven with an in-plane magnetic fields across sub-nanoscale defects, which eventually can trigger pinning, utilizing 
spin-polarized scanning tunneling microscopy~\cite{Holl2020}.
Skyrmions are deflected away from materials inhomogeneities with a skyrmion Hall angle proportional to the skyrmion velocity~\cite{nagaosa_topological_2013}.
Hence, if one wishes to use skyrmions as a mean to store and transport information it is imperative to understand how these textures interact with real defects as those that would be encountered in potential applications. Several works have contributed to fill this gap via model approaches mostly based on micromagnetism, due to the difficulty of studying isolated defects both in experimental and theoretical approaches. For instance, thorough phenomoenology-based investigations on current-induced dynamics have been performed by various groups. Pinning, depinning and creep-like motion was addressed in Refs.~\cite{Liu2013,Lin2013,Zhang2013,Mueller2015,Martinez2016,Navau2016,psaroudaki_quantum_2017} while the impact of disorder is the topic of Ref.~\cite{Reichhardt2015}.

From the ab-initio perspective, impact of single atomic defects on the stability of skyrmions was addressed in Ref.~\cite{Fernandes2018}, which showed that  he skyrmion-defect interaction profiles follow a universal pattern similar to what is known for cohesive, formation and surface energies. These defects enable spin-mixing magnetoresistances~\cite{crum_perpendicular_2015,Hanneken2015} of a new kind~\cite{Fernandes2019}  permitting and enhancing the all-electrical detection of non-collinear magnetic textures such as skyrmions. Going further, building multi-atomic defects permits the engineering of the energy landscape of skyrmions~\cite{Arjana2020}. For instance two repulsive atoms placed close to each other can give rise to a pinning defect.

In the present work the effect of $3d$ and $4d$ atomic defects on the motion of skyrmions in Pd/Fe/Ir(111) is studied using atomistic spin dynamics. This substrate is well characterised and known to host few nanometers-wide magnetic skyrmions~\cite{Romming2013,Romming2015,dupe_tailoring_2014,Simon2014,Leonov2016,Dias2016} stabilized by the presence of DMI. The influence of defects is considered using a parametrized Heisenberg Hamiltonian from first principles calculations performed in an earlier work~\cite{Fernandes2018}. Using this unique combination of techniques it is possible to address and describe systematically  the implications that these imperfections have on both the static and dynamical properties of skyrmions, with material dependent parameters. Thus, allowing a deeper understanding of the role of defects in skyrmion dynamics, and how these can be used to create pathways were skyrmion motion is preferred (see e.g.~Refs.\cite{Mueller2017,Fernandes2018,stosic_pinning_2017,Arjana2020}),  detection~\cite{Fernandes2019} and injection areas. The advantage of the present approach is the capability to take into account the influence that defects of different chemical species can have over the texture dynamics. Even more, the skyrmions under scrutiny in this work are of a very reduced size, with a radius of $\sim 4$ nm, which makes them ideal for potential applications, in contrast to the larger skyrmions studied with micromagnetism, thus highlighting the importance of understanding the skyrmion-defect interaction for any possible application.

The present paper is organized in the following way, first in section~\ref{sec:ASD} the computational methods used to describe the skyrmion dynamics will be introduced. In section~\ref{sec:Statics} the effect that single defects have over the static properties of the magnetic texture. Afterwards, in section~\ref{sec:Single_dyn} the dynamics of skyrmions under spin polarized currents will be described, and the effect that the chemical nature of the defects have over the skyrmion motion will be showcased. In section~\ref{sec:Thiele}, an anlysis of the skyrmion dynamics based on the Thiele Equation~\cite{Thiele} is provided before proposing diagrams showing the richness of  the motion regimes of skyrmions at the vicinity of pinning or repulsive defects. In  
section.~\ref{sec:double_track}, an example of a device is presented, where a skyrmion is driven through a double track made of pinning defects. Lastly, a conclusion is presented.

\section{Computational Methods\label{sec:ASD}}
\begin{figure}
\centering
\includegraphics[width=\columnwidth]{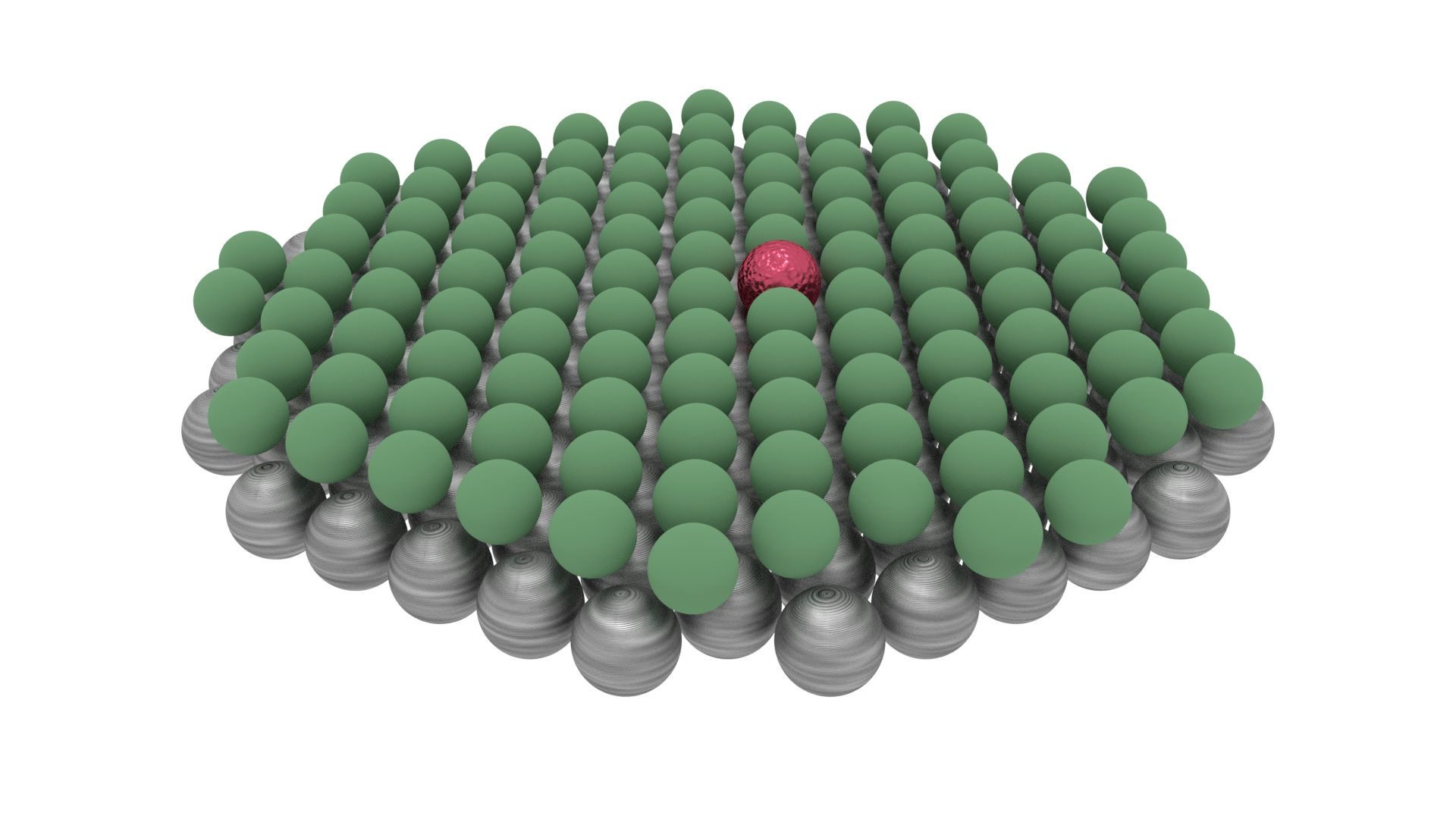}
\caption{Schematic view of the real space cluster  where the impurity is located and embedded in the perfect substrate. The green atoms indicate the Pd layer, the grey atoms the Fe layer and the red atom is the considered impurity. \label{fig:cluster}}
\end{figure}

The studied system is the a bilayer of Pd/Fe deposited on Ir(111), with single $3d$ (Cr,Mn,Co) and $4d$ (Nb,Tc,Ru) atomic defects embedded in the Pd layer (see Fig.~\ref{fig:cluster}). 

The calculations are based on density functional theory considering the local spin density approximation~\cite{Vosko1980}. We use the all-electron full-potential scalar-relativistic Korringa-Kohn-Rostoker (KKR) Green function method~\cite{Papanikolaou2002,Bauer2013} with spin-orbit coupling added self-consistently, which permits to embed single magnetic skyrmions and the defects in the magnetic substrate. The embedding technique allows one to obtain the effect of these single atomic defects via first principles calculations without the need of supercells and/or alloying techniques. Basic properties such as the magnetic moments $\mathbf{m}_i$ and the magnetocrystalline energy (MAE) are then obtained.

To extract the pairwise Heisenberg exchange parameters, $J_{ij}$, the Dzyaloshinskii-Moriya vectors, $\mathbf{D}_{ij}$, one single interation was performed using the infinitesimal rotation method\cite{Ebert2009,Liechtenstein198765} considering a k-mesh of $200 \times 200$ and an angular momentum cut-off at $l_\text{max} = 3$.
The Pd overlayer carries a sizable spin-moment of $\approx0.3 \mu_\text{B}$ induced by the Fe atoms, which is incorporated in the atomistic model via a renormalization scheme of the substrate's magnetic exchange interactions (MEI) as described in Ref.\cite{PhysRevB.82.214409}.  The magnetic states are investigated via an atomistic spin-dynamics approach~\cite{0953-8984-20-31-315203,eriksson2017atomistic} extended to treat an embedding problem similar to the {\it ab initio} method.

The obtained exchange interactions can then be used to describe the magnetic ground state and dynamics via the Landau-Lifshitz-Gilbert (LLG) equation of motion of the magnetic moment as implemented in the \texttt{UppASD} software package~\cite{0953-8984-20-31-315203,eriksson2017atomistic} : 
\begin{equation}
 \frac{d \mathbf{m}_i}{d t}=-\frac{\gamma}{1+\alpha^2}\left[\mathbf{m}_i \times \mathbf{B}_\text{eff}^i+ \frac{\alpha}{m_i}\mathbf{m}_i \times \left( \mathbf{m}_i \times \mathbf{B}_\text{eff}^i\right)\right]\,.
 \label{eq:LL}
\end{equation}
Each atomic moment, $\mathbf{m}_i$, is considered to be a three dimensional (3D) vector with constant magnitude. $\gamma$ is the gyromagnetic ratio, $\alpha$ is the Gilbert damping controlling the dissipation of angular momentum and energy from the magnetic subsystem (assumed here to be equal to 0.05) and $\mathbf{B}_\text{eff}^i$ is the effective field acting over the $i$-th site. The effective field considered here has the following form
\begin{equation}
 \mathbf{B}^{eff}_i=-\frac{\partial \mathcal{H}}{\partial \mathbf{m}_i}+\mathbf{b}_i(T)
\end{equation}
where, $\mathcal{H}$ being and extended Heisenberg Hamiltonian describing the system, and  $\mathbf{b}_i(T)$ is an stochastic field is included in order to account for temperature effects by using Langevin dynamics~\cite{0953-8984-20-31-315203}. 

In the following work the considered extended Heisenberg Hamiltonian is the following
\begin{equation}
\begin{split}
\mathcal{H}=&-\frac{1}{2}\sum_{i,j}J_{ij}\mathbf{m}_i\cdot\mathbf{m}_j -\sum_{i,j}\mathbf{D}_{ij}\cdot\left(\mathbf{m}_i\times\mathbf{m}_j\right)\\&-\sum_i K_i\left(\hat{e}_K^i\cdot\hat{m}_i\right)-\sum_{i}\mathbf{m}_i\cdot\mathbf{B}_\text{ext}
\end{split}
\label{eq:Hamiltonian}
\end{equation}
where $K_i$ is the magnitude of the anisotropy constant at the $i$-th atomic site, $\hat{e}_K^i$ indicates the anisotropy axis and $\mathbf{B}_\text{ext}$ is an applied external field.

This thermal magnetic field $\mathbf{b}_i(t)$, has a Gaussian shape and its average value in time is zero $\langle \mathbf{b}_i(t) \rangle=0$. It also exhibits no correlation in space or between it's different components:
\begin{equation}
\langle b_i^k(t) b_j^l(t') \rangle=2D\delta_{ij}\delta_{kl}\delta(t-t')
\end{equation}
with $D$ being the amplitude of the field.  $\displaystyle D=\frac{\alpha}{(1+\alpha^2)}\frac{k_B T}{\mu_{B} m}$, $i$ and $j$ are the lattice sites, $k$ and $l$ are the vector coordinates, $k_B$ is the Boltzmann constant, $T$ is the temperature, assumed here to be equal to 0.001 K, $\mu_{B}$ is the Bohr magneton and $\left|\mathbf{m}\right|$ is the magnitude of the magnetic moment.

The effect of spin polarized currents along the $x$-direction is taken into account by adding the adiabatic and non-adiabatic torques resulting from the STT~\cite{Li2004} to Eq.~\ref{eq:LL}
\begin{align}
\frac{\text{d}\mathbf{m}_i}{\text{d}t}=&-\frac{\gamma}{1+\alpha^2}\; \mathbf{m}_i\times
\left[ \mathbf{B}_\text{eff}^i+\frac{\alpha}{m_i}\left(\mathbf{m}_i\times\mathbf{B}_\text{eff}^i\right)\right]\nonumber\\
&+\frac{1+\beta\alpha}{1+\alpha^2}\frac{{u}_x}{m_i^2}\;\mathbf{m}_i\times\left(\mathbf{m}_i\times\frac{\partial \mathbf{m}_i}{\partial x}\right)\nonumber\\
&-\frac{\alpha-\beta}{1+\alpha^2}\frac{{u}_x}{m_i}\;\mathbf{m}_i\times\frac{\partial \mathbf{m}_i}{\partial x}
\label{eq:LLG_STT_first}
\end{align}
where, $\beta$ is the non-adiabatic parameter and $\mathbf{u}$ is related to the applied current density, $\mathbf{j}_e$, by $\mathbf{u}=\frac{\mathbf{j_e}Pg\mu_B}{2eM_s}$. $P$ is the polarization, $g$ the Land\'e g-factor, $e$ the electronic charge, and $M_s$ the saturation magnetization of the system.

To better investigate the impact of defects on the motion of skyrmions, we assumed $\beta = \alpha$ cancelling thereby the third term of the right side of Eq.~\ref{eq:LLG_STT_first}. This permits the motion of the skyrmion in the defect-free region along the direction of the applied current. Therefore, the equation of motion that is solved in practice reads:
\begin{align}
\frac{\text{d}\mathbf{m}_i}{\text{d}t}=&-\frac{\gamma}{1+\alpha^2}\mathbf{m}_i\times 
\left[ \mathbf{B}_\text{eff}^i+\frac{\alpha}{m_i} \left(\mathbf{m}_i\times\mathbf{B}_\text{eff}^i\right)\right] \nonumber\\
&+\frac{{u}_x}{ m_i^2}\mathbf{m}_i\times \left(\mathbf{m}_i\times {\frac{\partial \mathbf{m}_i}{\partial x}}\right).
\label{eq:LLG_STT}
\end{align}

\begin{figure}
\centering
\includegraphics[width=\columnwidth]{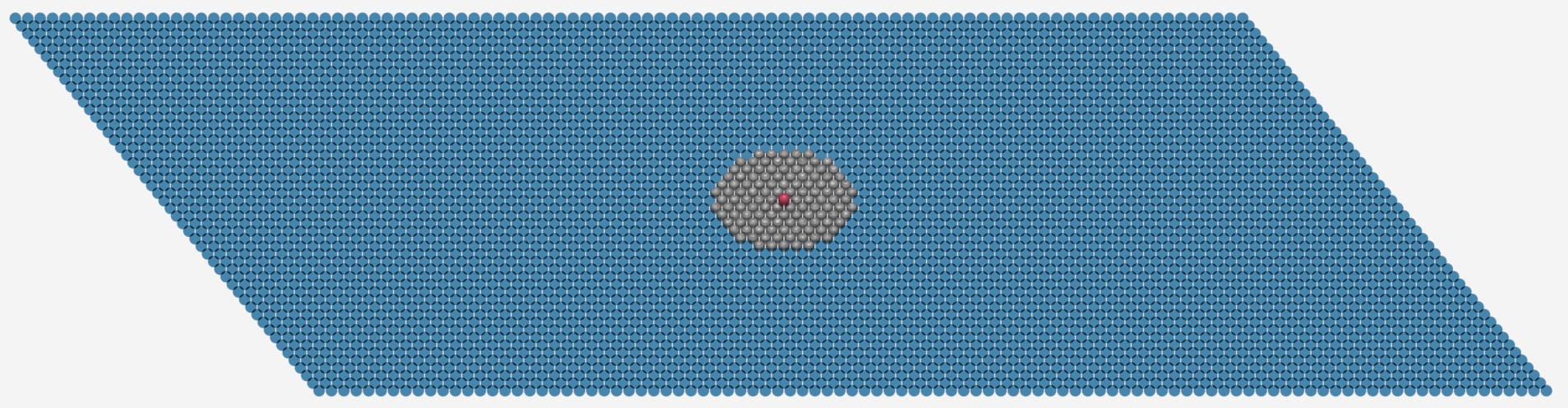}
\caption{Schematic visualization of of the impurity cluster embedded in the perfect substrate, as assumed in the atomistic spin dynamics simulations. The blue region indicates the magnetic moments with parameters obtained from the clean system,  the grey shaded region contains the atoms defining the impurity cluster while the red atom indicates the defect.\label{fig:ASD_cluster}}
\end{figure}

The treatment of the defects is performed by considering a simulation box in which the Hamiltonian for each atom is parameterized by the values corresponding to the clean system, except in the \textquotedblleft impurity cluster\textquotedblright\ region, as schematically shown in Fig.~\ref{fig:ASD_cluster}.  In this area the parameters are given by those obtained from the first principles simulations with defects, leading to local changes of the exchange couplings, $J_{ij}$, Dzyaloshinskii-Moriya vectors, $\mathbf{D}_{ij}$, anisotropy constant $K_i$ and magnitude of the magnetic moments $\mathbf{m}_i$.

\section{Results\label{sec:Results}}

In the following section, we address the results obtained  from atomistic spin dynamics using ab-initio parameters for skyrmions in Pd/Fe/Ir(111). The first aspect that must be studied in this treatment is the stability of the magnetic textures at the vicinity of the investigated atomic defects. 

\subsection{Static properties\label{sec:Statics}}

In the work from Lima Fernandes et al.~\cite{Fernandes2018} the energetics of magnetic skyrmions in Pd/Fe/Ir(111) was studied via full ab-initio calculations. In that work,  3$d$ impurities were found mostly of repulsive nature in contrast to mostly pinning 4d impurities. This was explained in terms of the hybridization of electronic states of the defects with those of the substrate, which an be translated to a competition of mechanisms affecting: (i) the magnetic interaction of the impurities with the Fe substrate and (ii) the modification induced by the presence of defects on the surface magnetic interactions.  On the one hand, the defects tend to decrease the magnetic interactions among the neighboring Fe atoms of the substrate. This leads to pinning since it favors non-collinearity among the substrate spin moments and stabilize skyrmions at the vicinity of the defects. On the other hand, the magnetic interaction between the defects and the Fe substrate  provides an additional magnetic exchange interaction, which tends to stiffen the surface magnetization disfavoring the presence of a magnetic skyrmion resulting in a repulsion behavior.

However, in those first-principles calculations the effect that an external magnetic field has over the skyrmion profile was not investigated. This is addressed in the current work using the aforementioned  ab-initio parameterized extended Heisenberg model. 

In order to determine the effect of  different impurities on the skyrmion radius, a skyrmion profile was stabilized with its center being located at the vicinity of a impurity site, in a simulation box of 100x100 repetitions of the unit cell. As some of the defects can be repulsive, the spin at the skyrmion core was fixed, whilst the rest of the magnetic moments in the sample was allowed to evolve under the LLG equation. An external magnetic field, $\mathbf{B}_\text{ext}$, was applied along the +z direction, that is  with an anti-parallel orientation with respect to the spins at the skyrmion core. The field is needed to deform the spin spiral ground state to form isolated skyrmions in a ferromagnetic background. The magnitude of the external field was varied between $5-16$ T. When $B_\text{ext}=4$ T, the system is in the region where helical spin spirals and skyrmions are very close together in energy~\cite{dupe_tailoring_2014}.

\begin{figure}
\centering
\includegraphics[width=\columnwidth]{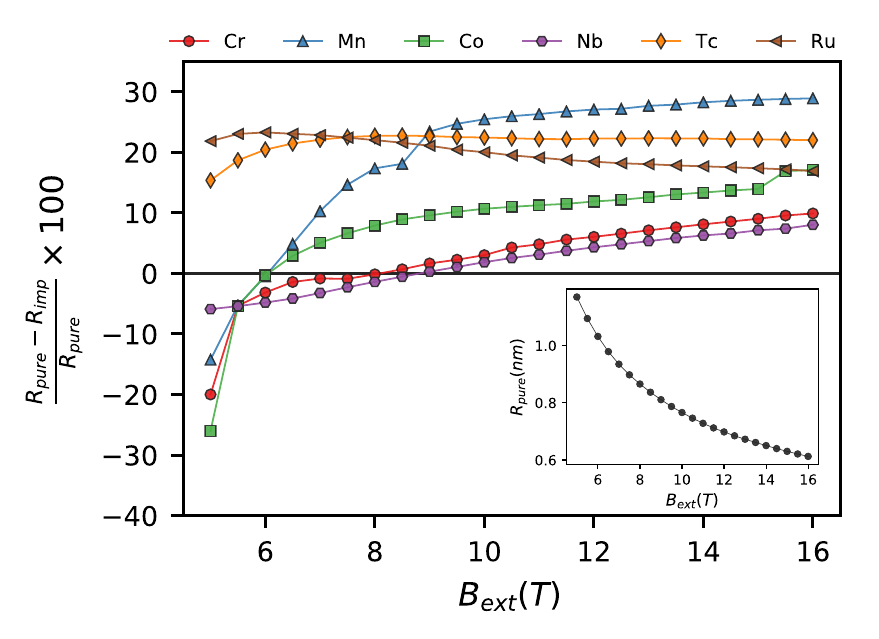}
\caption{Percentage difference between the skyrmion radius in the clean sample  and the one with a defect as function of an applied external magnetic field $\mathbf{B}_\text{ext}$.  Positive (negative) values indicate a shrinking (increase) of the skyrmion. The inset illustrates the field dependence of the skyrmion radius in the defect-free substrate.\label{fig:Skyr_radius}}
\end{figure}

Fig.~\ref{fig:Skyr_radius} illustrates the dependence of the relative change of the texture size as function of magnetic field for the different types of investigated impurities. The inset shows the  monotonous decrease of the skyrmion radius in the defect-free surface when  increasing the magnitude of the field.  However, and perhaps most interestingly, the size of the skyrmion strongly depends  on the impurities chemical nature. 
The $4d$ defects Tc and Ru, which are of pinning nature decrease the radius of the skyrmion for the whole range of investigated magnetic field. The concomitant defect-induced reduction of the substrate MEI and the weak defect-substrate MEI makes the skyrmion more maleable and therefore the magnetic field has the ability to reduce skyrmion size.
While the relative change of the skyrmion size is rather constant as function of the field, the repulsive 4$d$ and 3$d$ impurities, Nb, Cr, Mn and Co, decrease the radius of the skyrmion for low fields before triggering a size increase at fields ranging from about 6 to 8 T. 
Since the core of the skyrmion is fixed at the vicinity of the defects while experiencing repulsion, the skyrmion tries to move away, which enforces a deformation extending its size. This is possible for small enough magnetic field. In the latter case, stark elliptical deformations (see coming discussion) occur, which are favored by the small energy difference of the helical spin spiral and of the skyrmion. Large magnetic fields shrink dramatically the skyrmions reducing thereby their internal distortion. These results showcase the possibility of tuning the size of skyrmion owing to the presence of defects.

\begin{figure}
\centering
\includegraphics[width=1.\columnwidth]{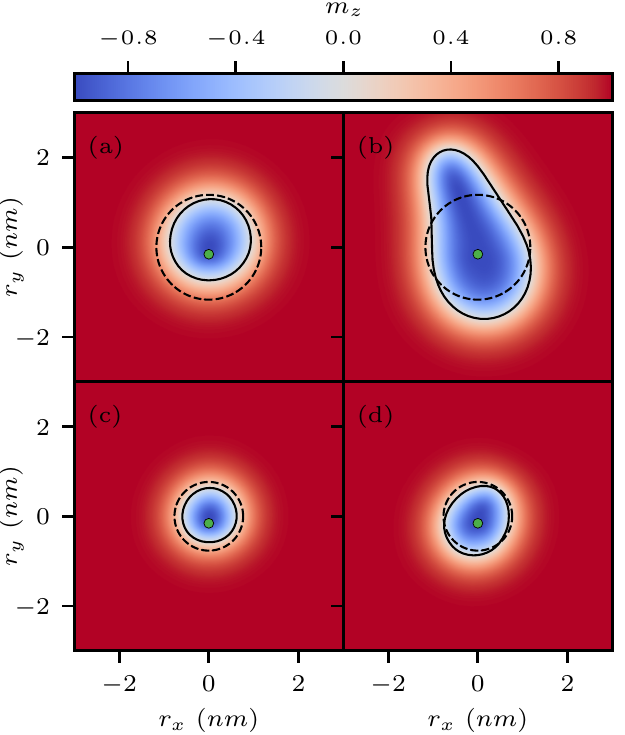}
\caption{Skyrmion profiles (full line) at the closest vicinity to Ru (left) and Mn (right) impurities. The position of the impurity is indicated with the green circle while the dashed line represent the skyrmion profile in the defect-free region. The color scale correspond to the out-of-plane component of the atomic magnetic moment, $m_z$, in the Fe monolayer. For an external magnetic field of 5 T, the skyrmion shrinks close to Ru (a) but increases elliptically in size when located underneath Mn (b). A larger magnetic field of 10 T restores more or less the isotropy of the skyrmion, which experience a size decrease in both cases: close to Ru (c) and Mn (d).   \label{fig:Skyr_deformation}}
\end{figure}

The impact of the impurities on the skyrmion profile is illustrated in Fig.~\ref{fig:Skyr_deformation} for two examples: the pinning Ru and repulsive Mn defects. Full lines indicate the deformed skyrmion while dashed lines represent the original skyrmion in the defect-free region. For an external magnetic field of $B_\text{ext}=5$ T, it can be seen that, on the one hand, the Ru impurity, Fig.~\ref{fig:Skyr_deformation}\textbf{(a)}, shrinks the skyrmion (by about 20\% as indicated in Fig.~\ref{fig:Skyr_radius}) distorted. We note moreover that the distortion is asymmetric, that is the perturbation goes beyond merely changing the radius of the skyrmion. On the other hand the case of  Mn impurity, Fig.~\ref{fig:Skyr_deformation}\textbf{(b)}, is of special interest, since the large increase of the skyrmion size, by about as 15\% observed in Fig.~\ref{fig:Skyr_radius}, results from the large deformation of the texture profile.  
As the magnitude of the external magnetic field increases to $B_\text{ext}=10$ T, the relative change of the skymrion size at the vicinity of Ru is similar to what was obtained for the weaker field, see Fig.~\ref{fig:Skyr_deformation}\textbf{(c)}. The change is however extreme for Mn, for which a quasi-isotropic skyrmion shape is restored as shown in Fig.~\ref{fig:Skyr_deformation}\textbf{(d)}.

\subsection{Skyrmion dynamics \label{sec:Single_dyn}}

We now study the dynamics associated with single magnetic skyrmions driven towards point-defects after applying a current along the $x$-direction (see Eq.~\ref{eq:LLG_STT}). For conciseness, we focus our following analyses on single 3$d$ (Co, Cr) and 4$d$ (Tc, Ru) atomic defect considering an external magnetic field of 10 T.

\begin{figure}[t!]
\centering
\includegraphics[width=1.\columnwidth]{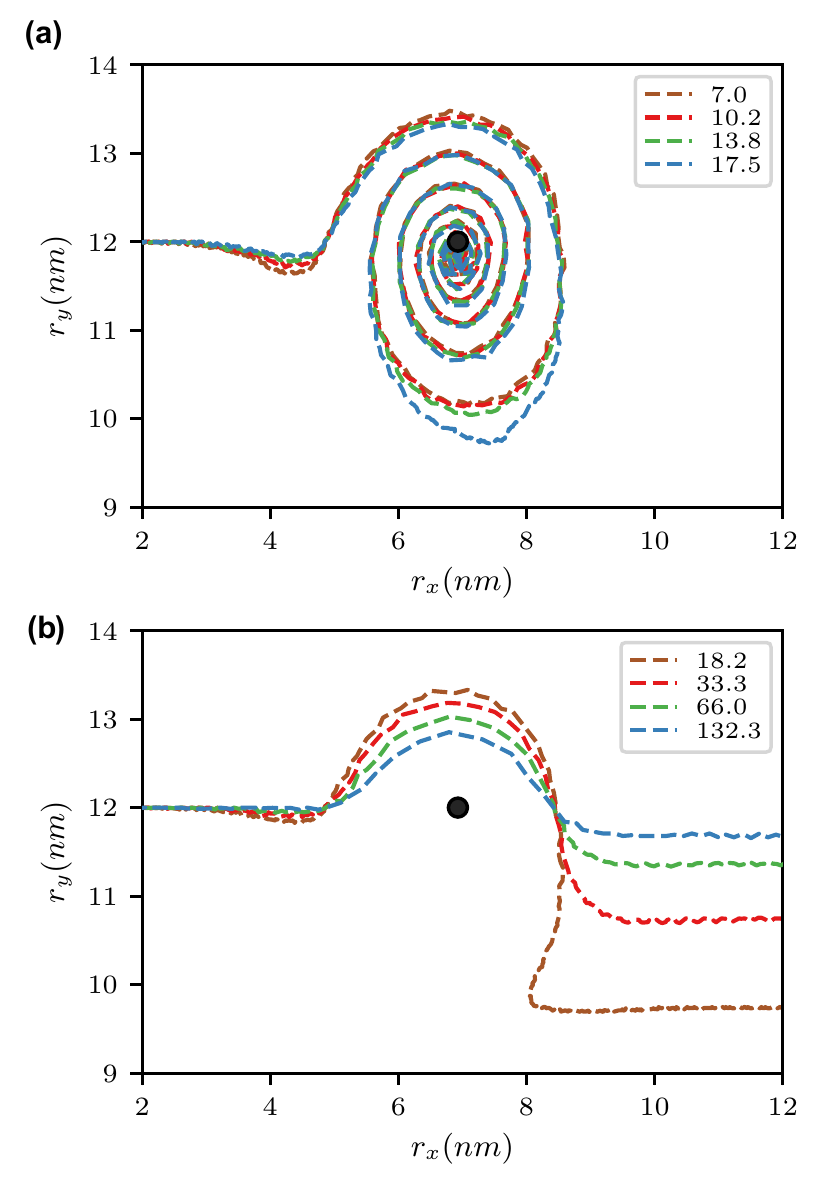}
\caption{Trajectories of a skyrmion stabilized with an external magnetic field  $B_\text{ext}=10$ T driven with different incoming velocities (m/s) along the $x$-direction towards the pinning Ru defect. The position of the latter is indicated with a black circle. The skyrmion gets either pinned (a) or deflected (b) depending on the intensity of the current density. Pinning occurs after a clockwise gyrotropic motion. If the current is strong enough, the skyrmion is deflected and the skyrmion escapes the  impurity's pinning force field but experiencing a shift in the $y$-coordinate. \label{fig:skyr_trajectories_Ru}}
\end{figure}

\begin{figure}[t!]
\centering
\includegraphics[width=1.\columnwidth]{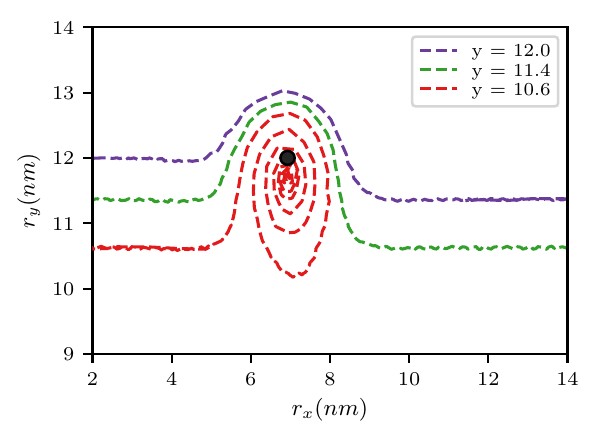}
\caption{Skyrmion trajectory around the pinning Ru impurity as function of the incoming coordinate. The latter dictates the skyrmion pinning or defect avoidance. An effective incoming velocity of 66.0 m/s is assumed with an external magnetic field of 10 T. 
The position of the impurity is indicated with the black circle.  \label{fig:Skyr_trajectories_change-y}}
\end{figure}

\textbf{Pinning defects}. We show in Fig.~\ref{fig:skyr_trajectories_Ru} the trajectories of the skyrmion obtained after applying different current  intensities along the $x$-direction. This is made possible thanks to the STT encoded in the 
the third term of Eq.~\ref{eq:LLG_STT} which drives the skyrmion towards a Ru impurity. The effective incoming skyrmion velocity in m/s is listed, which is the velocity of the skyrmion in the defect-free region. Note that a similar behavior characterizes the case of Tc. 

Since Ru is of  pinning nature, we expect the impurity to attract the passing skymion and pin it. If the current is not strong enough, as illustrated in  Fig.~\ref{fig:skyr_trajectories_Ru}(a), the skyrmion gravitates around Ru before getting pinned following a clockwise spiraling trajectory. If the incoming skyrmion velocity is larger than $\SI{18}{\meter/\second}$, the skyrmion escapes the pinning forces induced by the impurity (see Fig.~\ref{fig:skyr_trajectories_Ru}(b)). The larger the current is, the closer the skyrmion is passing by the defect. Interestingly the skyrmion experiences a shift to a lower $y$-coordinate, named henceforth $y$-deflection, and then follows a straight line along the direction of the applied current. 

\begin{figure}
\centering
\includegraphics[width=1.\columnwidth]{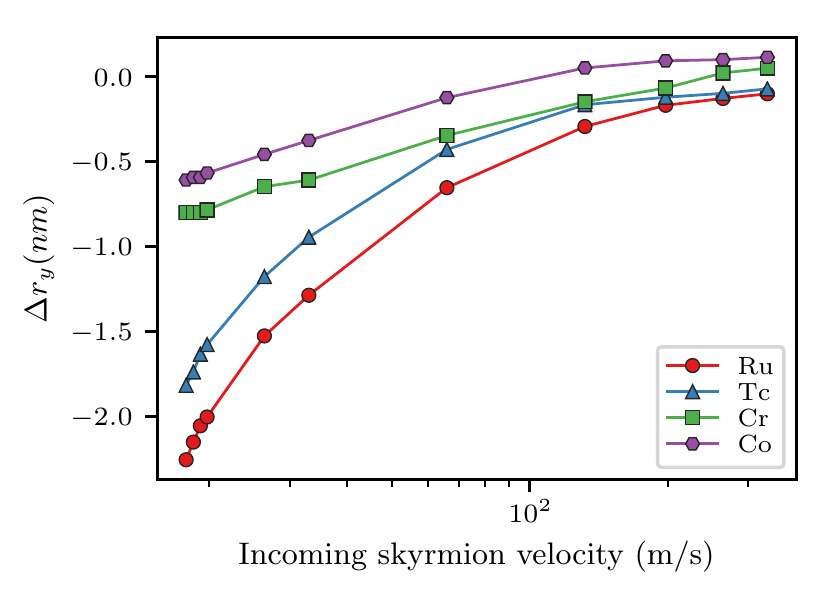}
\caption{$y$-deflection, i.e. change of the $y$-coordinate of the skyrmion, after leaving the area containing the defect as function of the incoming skyrmion velocity. A magnetic field of 10 T is applied. \label{fig:y-deflection_current}}
\end{figure}

As demonstrated in  Fig.~\ref{fig:Skyr_trajectories_change-y}, the escape-coordinate can be manipulated by modifying the incoming location of the skyrmion. The same figure shows that this  impacts the escape ability of the skyrmion, which  can even get pinned. Note that the incoming skyrmion velocity of $\SI{66}{\meter/\second}$ is strong enough to make the skymion bypass the defect (green line in Fig.~\ref{fig:skyr_trajectories_Ru}(b)). This behavior is similar to that of the other investigated defects, as shown in 
Fig.~\ref{fig:y-deflection_current} in a logarithmic-scale as function of the incoming skyrmion velocity. In general, the faster the skyrmion is, the less it feels the defect.  

\begin{figure}
\centering
\includegraphics[width=1.\columnwidth]{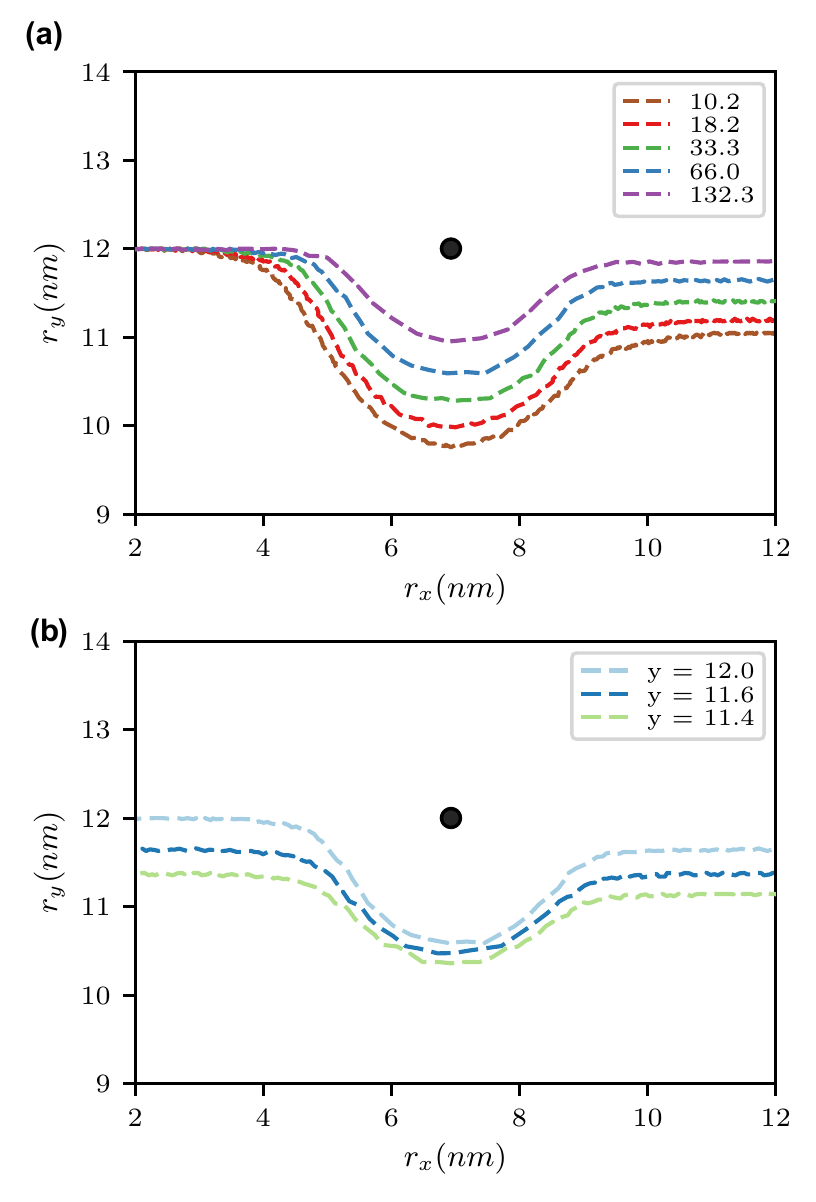}
\caption{Trajectories of a skyrmion around the repulsive Cr defect 
for different incoming velocities (a) and considering different incoming coordinates (b), where we assumed an incoming skyrmion velocity of 66 m/s. The position of the impurity is indicated with the black circle.\label{fig:skyr_traj_Cr}}
\end{figure}

\textbf{Repulsive defects}. Surprisingly, when considering repulsive impurities such as Cr, shown in  Fig.~\ref{fig:skyr_traj_Cr}(a), or Co, the path followed by the skyrmion seems similar to that experienced when directed towards a pinning defect. After the initial impurity avoidance, the skyrmion seems to gravitates around the impurity for  almost half a period before the escape. The circulation around the Cr indicates that there is a sort of pinning path emerging when the skyrmion is brought to a dynamical state. Even more intriguing is that the deflection of the defect is counter clockwise, i.e. in the opposite direction than what was observed for the pinning defects. Moreover, increasing the incoming skyrmion velocity decreases the skyrmion-defect separation as found for the pinning defects. Similarly to the latter impurities, $y$-deflection occurs for repulsive impurities but with a magnitude that is about a factor of two smaller (see Figs.~\ref{fig:skyr_traj_Cr}(b)  and~\ref{fig:y-deflection_current}). 

\subsection{Analysis of the dynamics with the Thiele equation\label{sec:Thiele}}
\begin{figure}
\centering
\includegraphics[width=1.00\columnwidth]{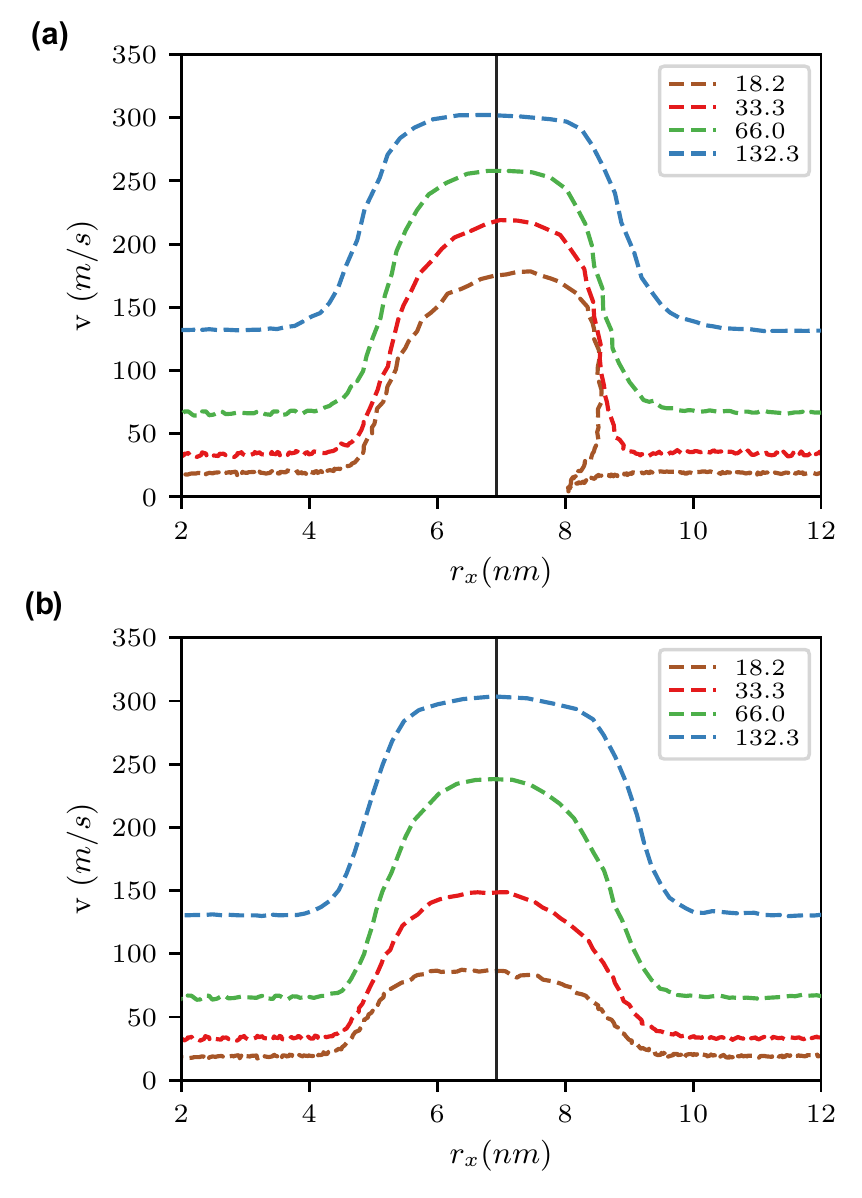}
\caption{Velocity of the skyrmions   as function of its position under different applied currents. The case of a pinning defect (Ru) is shown in (a) while the repulsive one (Cr) is in (b) with the position of the impurity indicated by the perpendicular line. The external magnetic field is set to $B_\text{ext}=10$ T. \label{fig:skyr_velocity_Ru}}
\end{figure}

\begin{figure*}[ht!]
\centering
\includegraphics[width=2.00\columnwidth]{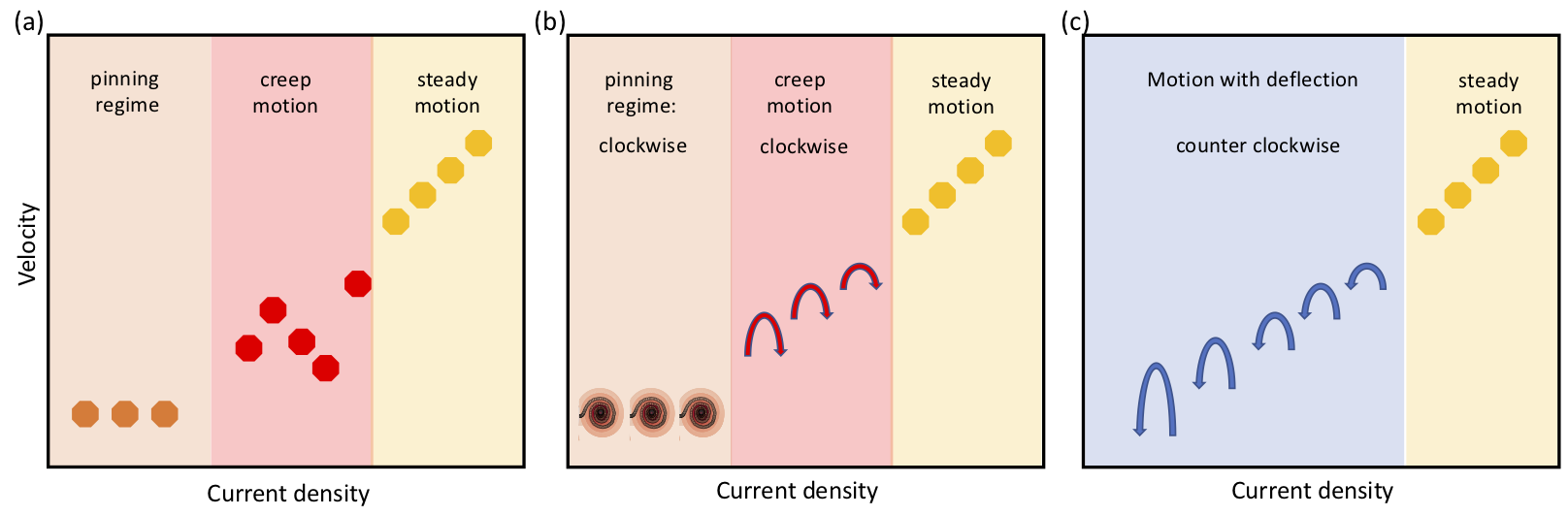}
\caption{Schematic diagrams for the skyrmion velocity as function of the applied current density. (a) Illustration of the usual picture known for skyrmions moving across pinning defects, where three regimes are defined: pinning, creep motion and steady motion. If the current is not strong enough the skyrmion is pinned but if it moves it can still be pinned depending on the pinning force of the impurities. At large enough currents, the skyrmion moves steadily. (b) Our results indicate that for attractive defects, the pinning regime is dynamically rich since the skyrmions follow a gyrotropic clockwise spiraling motion. The skyrmion velocity increases at the vicinity of the defects but following a clockwise motion indicated by the arrows. (c) Repulsive defects induce also a rich dynamical behavior where the skyrmion deflection occurs in a counter clockwise fashion with an enhanced skyrmion velocity.    \label{fig:schematic_velocity}}
\end{figure*}

Here we explain the various observed phenomena  invoking the Thiele equation~\cite{Thiele}, which assumes rigid motion of non-collinear smoothly and weakly rotating spin-textures~\cite{Zhang2013,Lin2013}:
\begin{eqnarray}
\boldsymbol{G} \times (\boldsymbol{u} - \boldsymbol{v} ) + \alpha D(\boldsymbol{u} - \boldsymbol{v} ) = \boldsymbol{F}_d \,  , 
\end{eqnarray}
where we considered $\beta = \alpha$. 
The first term is the Magnus force, where $G = 4\pi Q  \hat{z}$ is the gyrovector responsible for the skyrmion Hall effect whenever the topological charge $Q$ is finite. $Q = -1$ for the studied skyrmion. This leads to the deflection of the skyrmion from the direction of the applied current.  The second term is dissipative and $\boldsymbol{F}_d$ is the force induced by the defect. The force is isotropic and can be written as
\begin{eqnarray}
\boldsymbol{F}_d (\boldsymbol{R}) = \mp f(R) (\cos(\theta),\sin(\theta)),
\end{eqnarray}
where the minus (plus) sign indicates the pinning (repulsive) nature of the defect. $\boldsymbol{R}$ is the vector connecting the skyrmion center of mass and the defect position. In the chosen convention, the skyrmion arrives at an angle $\theta = \pi$.

$\boldsymbol{u}$ is related to the applied current density (see Eq.~\ref{eq:LLG_STT_first}) pointing along $-\hat{x}$ while $\boldsymbol{v}$ is the velocity of the skyrmion, whose components are:
\begin{eqnarray}
v_x &= &\frac{1}{\alpha^2D^2 + 16\pi^2Q^2} (\alpha D F_d^x + 4\pi Q F_d^y) + |u_x|\\
v_y &= & \frac{1}{\alpha^2D^2 + 16\pi^2 Q^2} (-4\pi Q F_d^x + \alpha D F_d^y  ).
\end{eqnarray}

\begin{figure}[ht!]
\centering
\includegraphics[width=.9\columnwidth]{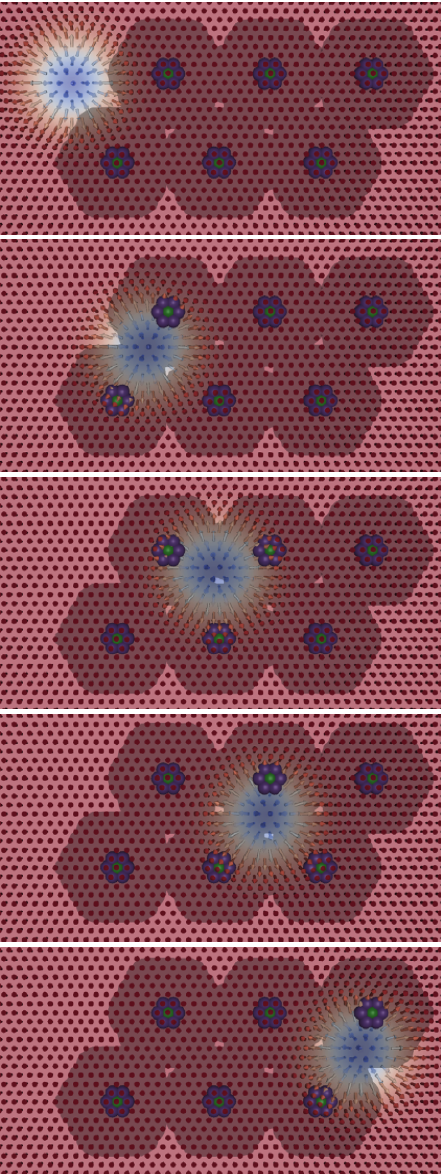}
\caption{Snapshots of the motion of a skyrmion in a double track made of pinning Tc defects. The skyrmion comes from the left (upper figure) with an incoming velocity of 66 m/s. Engineering of the inter-impurity distances allows controlling the skyrmion deflection permitting the control of the skyrmion trajectory. The dark clusters illustrate the range of the defect-substrate magnetic exchange interactions. \label{fig:double_track}}
\end{figure}

These equations indicate that in the defect-free region, the skyrmion would move along the $x$-direction with the velocity $|u_x|$. As soon as it feels the force emanating by the defect, deflection occurs independently from the nature of the defect being pinning or repulsive. The latter defines the direction followed by the skyrmion. A pinning (repulsive) defect leads to a clockwise (counter clockwise) path, which can be deduced from the previous equations. If the skyrmion, assuming $Q = -1$, and the defect were at the same $y$-coordinate and for $\theta = \pi$, $F_d^y = 0 $:
\begin{eqnarray}
v_y &=&  \frac{4\pi}{\alpha^2D^2 + 16\pi^2}  F_d^x = \pm \frac{4\pi}{\alpha^2D^2 + 16\pi}  f \,,
\end{eqnarray}
which means that a pinning (repulsive) defect induces a finite $y$-component of the velocity that is positive (negative) explaining the clockwise (anticlockwise) motion.

While it is expected to observe the circular motion of the skyrmion around the pinning defect, this is not the case for the repulsive defect. This gyromagnetic motion indicates that at $\theta = \pi/2$ (or close to that value), the $y$-component of the velocity should cancel out leading to 
\begin{eqnarray}
4\pi F_d^x &=& - \alpha D F_d^y \, \ \ \text{and} \,  \ \ \cos(\theta) = \frac{\alpha D}{4\pi} \sin(\theta).
\end{eqnarray}
This shows that a rotation of the skyrmion is expected for both kind of impurities at an angle $\theta = {\pi}/{2}  + \frac{\alpha D}{4\pi}$ for a pinning defect or ${3\pi}/{2}  + \frac{\alpha D}{4\pi}$ for the repulsive one. In other words, the maximum (minimum) of the $y$-coordinate of the skyrmion occurs at a position slightly shifted to the right-side (left-side) of the defect if it is pinning (repulsive), as observed in Figs.~\ref{fig:skyr_trajectories_Ru} and~\ref{fig:skyr_traj_Cr}.

From the trajectory followed by the skyrmion, one can identify whether the defect is pinning or repelling. Both trajectories are similar and it is reasonable to ask, whether dynamical effect could lead to a spiraling path for more than half a period around the repulsive defect. 
This does not occur in our simulations, which can also be demonstrated from the Thiele equation. To realize the aforementioned  circulation, the $x$-component of the velocity has to cancel out at $\theta = 0$ where $F_d^y = 0$:
\begin{eqnarray}
\frac{\alpha D}{\alpha^2 D^2 + 16 \pi^2} F_d^x = - |u_x|.
\end{eqnarray}
This condition can be satisfied for the pinning impurity because of the direction of its force, as expected, but not for the repulsive defect.

The magnitude of the velocity changes in magnitude because of the presence of the impurities:
\begin{eqnarray}
v^2 &=& \frac{F^2}{\alpha^2 D^2 + 16 \pi^2} + u_x^2 \nonumber \\
&&+ 2\frac{ |u_x|}{\alpha^2 D^2 + 16 \pi^2} (\alpha D F_d^x - 4\pi F_d^y), \label{eq:velocity}
\end{eqnarray}
which shows that besides the expected impact of the applied current, the magnitude of the change depends on the nature of the defects. The latter is encoded in the first and third terms of the r.h.s of Eq.~\ref{eq:velocity}. 
The maximum change of the velocity is positive and expected around $\theta = \pi /2$ ($+ \pi$ for the repulsive defect) since 
\begin{eqnarray}
\alpha D F_d^x - 4\pi F_d^y &\simeq& 4 \pi f.
\end{eqnarray}
Our simulations illustrated in  Fig.~\ref{fig:skyr_velocity_Ru} indicate that for Ru, which is pinning, the change in velocity (maximum - minimum) is rather constant for incoming skyrmion velocities larger than 18.2 m/s. This means that the skyrmion follows a path of a constant force. This is not the case for the velocity obtained around the repulsive defect. Here, the velocity increases monotonically with the applied current.

\subsection{Skyrmion motion regimes\label{sec:schematic_velocity}}
The motion of skyrmions, similarly to domain walls, is known to follow three regimes depicted schematically in Fig.~\ref{fig:schematic_velocity}(a). If the applied current is weak, the skyrmion is pinned, which defines the pinning regime. As some threshold, the skyrmion can escape the defect before getting trapped by another one. This defines the creep motion before moving steadily in what defines the last regime: steady motion. Our work on ultrasmall skyrmions demonstrate the richness of the aforementioned regimes. For instance, pinning defects, illustrated in Fig.~\ref{fig:schematic_velocity}(b), induce gyrotropic clockwise motion of the skyrmion before experiencing pinning in the first regime. In the second one, the skyrmion can escape but experiences deflection in a clockwise fashion with an enhanced effective velocity at the defect vicinity. The case of motion across repulsive defects deserves also a schematic diagram shown in Fig.~\ref{fig:schematic_velocity}(c). Instead of pinning and creep motion regimes, the repulsive defects generate a deflection motion regime, where the skyrmion scatters at the impurity which modifies its location and follows an counter clockwise path.

\subsection{Skyrmion motion in a double track of defects\label{sec:double_track}}
In Fig.\ref{fig:double_track}, we illustrate how the skyrmion deflection at the vicinity of defects can be engineered to craft a device allowing the control the skyrmion trajectory. We position pinning Tc defects to  form a double track such that they have a slight overap of their long-range magnetic interactions, shown as a dark clusters around the impurities. At the vicinity of the impurity, the skyrmion coming with an initial velocity of 66 m/s avoids the two first defects. Then it follows path defined by the edges of the aforementioned dark clusters and result from the deflection compromise of all the surrounding defects. Thus the motion is naturally not straight, but the applied current enforce an average motion along the $x$-axis. Since the velocity around the defects is increased, the double track allows to enhance the skyrmion velocity.

\section{Conclusion}
We investigated via a two-pronged approach the magnetodynamics of single magnetic skyrmions at the vicinity of 3$d$ and 4$d$ atomic defects embedded in the Pd layer deposited on Fe/Ir(111) surface. Starting with ab-initio simulations based on the all-electron relativistic full-potential Korringa-Kohn-Rostoker Green functions method, we extracted the tensor of magnetic interactions, which we use to solve the Landau-Lifshitz-Gilbert equation of motion of the magnetic moments under the influence of the  spin-transfert-torque. The defects have a strong impact on the size and shape of the skyrmion, which is strongly dependent on the external magnetic field. Elliptical deformations were observed at the vicinity of the repelling 3$d$ impurities. Upon application of a current, and assuming the non-adiabatic parameter $\beta$ to be equal to the Gilbert damping $\alpha$, the skyrmion flows along the direction of the current. Skyrmion scattering at defects is investigated assuming a magnetic field of 10 T, which makes the skyrmions ultrasmall. At the vicinity of the defect, deflection occurs leading to opposite direction of the gyomagnetic precession depending on the interaction nature of the impurities. The precession  is found  clockwise for a pinning force and counter clockwise with a repulsive force. This distinction permits the identification of the nature of the skyrmion-defect complex being repulsive of pinning by the analysis of the skyrmion trajectory. The skyrmion manages to always escape a repelling impurity, which is not the case for a pinning defect. For the latter case, a minimum depinning current is required. The various observed phenomena seem to be qualitatively reasonably described via the Thiele equation, which is based on the rigid-core approximation. This is due probably due to the small size of the investigated skyrmions. Going beyond that limit~\cite{Ritzmann2018} is an interesting matter of investigation, which we will pursue in the context of a realistic description of spin-dynamics of skyrmion-defect complexes. Finally, we illustrated the potential use of defects to control the motion of ultrasmall skyrmions along well defined directions and enhance their effective velocity. This requires the control of the position of the impurities as well as their mutual distances. 

\section*{Acknwoledgments}
We thank Jan Masell (M\"uller) for fruitful discussions. This work is supported by the European Research Council (ERC) under the European Union's Horizon 2020 research and innovation programme (ERC-consolidator grant 681405 — DYNASORE). We acknowledge the computing time granted by the JARA-HPC Vergabegremium and VSR commission on the supercomputer JURECA at Forschungszentrum Jülich.

\bibliography{References}{}

\end{document}